\documentclass[twocolumn,aps,prc,superscriptaddress,preprintnumbers,floatfix,nofootinbib,showpacs]{revtex4}
\usepackage{graphicx}
\usepackage{dcolumn,multirow}
\usepackage{amsmath}

\begin{document}

\title{Determination of shear forces inside the proton \\ }

\newcommand{\JLAB}{Thomas Jefferson National Accelerator Facility, Newport News, VA 23606, USA}
\newcommand {\CNF} {Center for Nuclear Femtography, SURA, 1201 New York Ave. NW, Washington, DC 20005, USA }
\newcommand{\GWU} {Department of Physics - The George Washington University
	725 21st Street, NW, Washington, DC 20052}

\author{V.D.~Burkert} 
\affiliation{\JLAB} 
\author{L.~Elouadrhiri} 
\affiliation{\JLAB} 
\affiliation{\CNF}
\author{F.X.~Girod} 
\affiliation{\CNF}
\affiliation{\GWU}  

\date{\today}

\begin{abstract}
We report on the first determination of the shear forces on quarks inside the proton from experimental data on deeply 
virtual Compton scattering. The maximum shear force of $40 \pm 20 $~MeV fm$^{-1}$ occurs near 0.6~fm from 
the proton center, indicating where confinement forces may be strongest. On the macroscopic scale of the earth 
surface, this force corresponds to the weight of a mass of $\approx 650$~kg. The shear forces in the proton 
reverse direction at $r \approx 0.45$~fm from the center.  
\end{abstract} 

\maketitle

Protons and neutrons, generally referred to as nucleons, are the fundamental building blocks of nuclei and 
make up nearly 100\% of the mass of ordinary matter in the universe. They are composed of elementary objects, 
quarks and gluons. The latter are the carrier of the strong force. The distribution of quarks inside the proton 
is governed by the strong interaction as described by the theory of strong interaction, Quantum-Chromo-Dynamics (QCD). 

The internal mass and energy distributions, the forces on the quarks, and the angular 
momentum distribution inside the proton are largely unknown. These properties are encoded in the 
proton's matrix element of the energy-momentum tensor~\cite{Kobzarev:1962wt,Pagels:1966zza} and are expressed
in gravitational form factors (GFFs)~\cite{Ji:1996ek}. For decades the GFFs were considered 
unmeasurable due to the extreme weakness of the gravitational interaction. More recent theoretical 
development however showed that the GFFs may be indirectly probed in deeply virtual Compton 
scattering (DVCS)~\cite{Ji:1996nm,Radyushkin:1996nd}. 
 DVCS allows for the extraction of the internal proton structure expressed in the generalized 
parton distributions (GPDs)~\cite{Mueller:1994,Ji:1996ek}, and are the basis for the exploration of its gravitational 
properties~\cite{Polyakov:2002yz}. This new direction of nucleon structure research has recently resulted in 
the first estimate of the pressure distribution inside the proton based on experimental data~\cite{Burkert:2018bqq}. 
In this paper we report the first results of the tangential (shear) forces and their 
spatial distribution inside the proton.   
  
The critical part is the $D$-term, which represents the last unknown global property of the proton which,  
until recently, has remained unconstrained. In previous work the corresponding form factor $D(t)$ was determined 
in a range of 4-momentum-squared $-t$, and used to study the pressure distribution inside the proton. In this letter 
we employ $D(t)$ to determine the forces on the quarks in the proton, using the DVCS process as an effective proxy of the gravitational interaction. We first summarize the steps involved in this process.
 \begin{figure}[th]
\hspace{-0.5cm} \includegraphics[width=1.0\columnwidth]{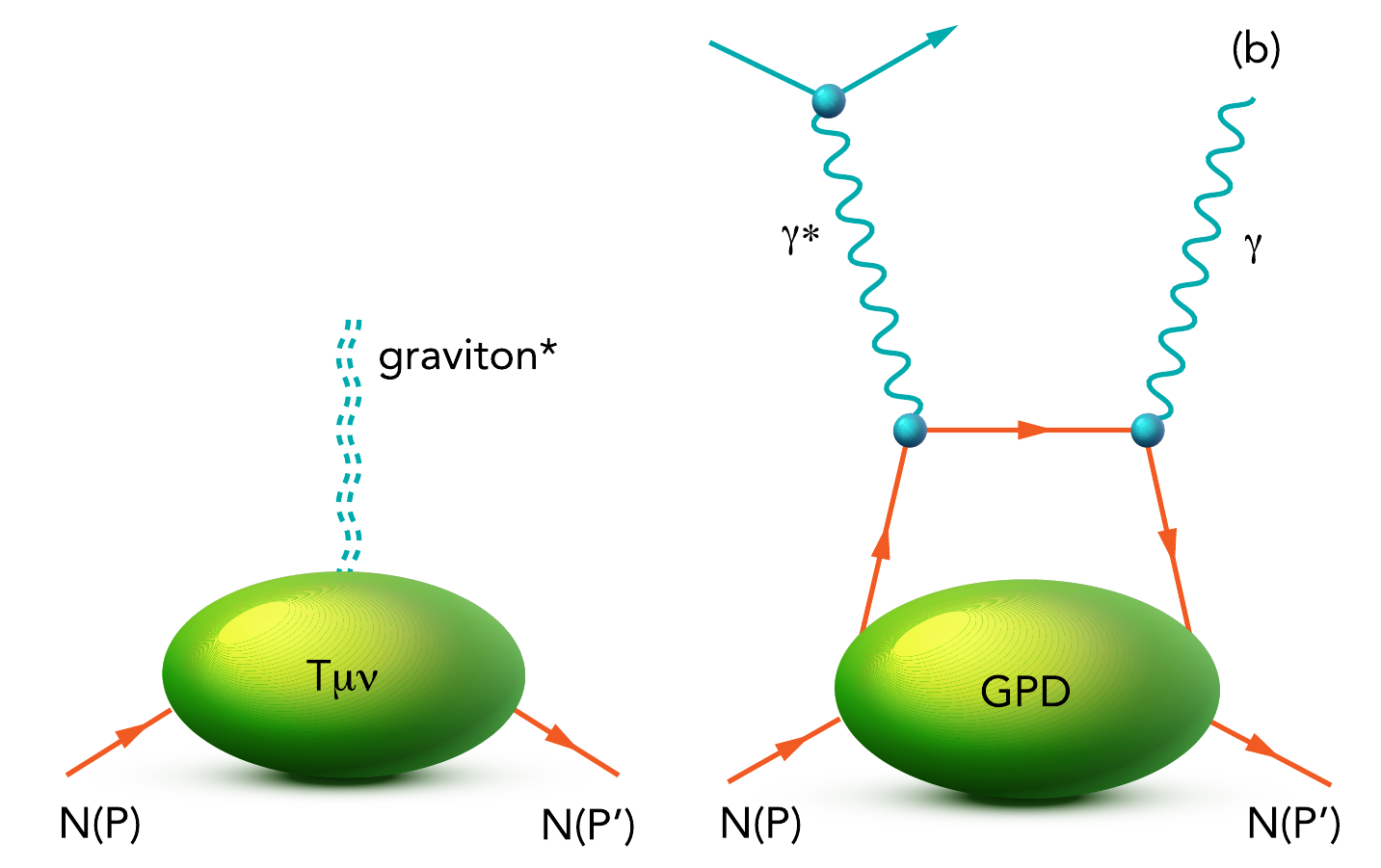}
\bigskip
\vspace{-0.5cm}\caption{Left graph: The graviton-proton interaction as a probe of the gravitational form factors. Gravity is a 
spin $J_G = 2$ tensor coupling. 
Right graph: The graviton-proton coupling is mimicked with the two $J_\gamma = 1$ photon vertices  in the leading 
handbag diagram in DVCS when integrating over the quark propagator. }
\label{dvcs-graviton}
\end{figure} 

The basic process is the handbag diagram shown in Fig.~\ref{dvcs-graviton} in leading twist approximation.  
All particles involved, the scattered electron, the emitted photon and the recoil proton are measured in coincidence 
to establish the exclusive DVCS process. The two high-energy photons each with spin $J_{\gamma} = 1$ that couple 
to the same quark in the proton, have the characteristics as a single graviton of spin $J_G = 2$ coupling to the quark, when integrated over the quark propagator. 
As the electromagnetic coupling to quarks is many orders of magnitude stronger than gravity, we can use the DVCS 
process as an effective proxy to study the gravitational properties of the proton experimentally. 

 The DVCS process is described, in leading twist, by 3 chiral even GPDs, of which $H(x,\xi,t)$ is most relevant in this 
 study, where $x$ is the 
 momentum fraction of the struck quark, $\xi$ is the longitudinal momentum fraction transferred to the struck quark, 
 and $t$ is the 4-momentum transfer to the proton.  At sufficiently high energies, the process factorizes into the coupling of the virtual 
 and real photon to the struck quark, and into the non-perturbative part described by the GPDs. For the proton, GPD $H$ dominates 
 the process, while other contributions are expected to be smaller, and kinematically suppressed. We neglect these smaller 
 contributions in the analysis and  include their estimated contributions into the systematic uncertainties of the final results.  

The GPD $H$ is directly mapped to the gravitational form factors $d_1(t)$ and $M_2(t)$ in a sum rule~\cite{Ji:1996nm} 
involving  the second Mellin moment of the GPD $H$:
\begin{eqnarray}
\int \mathrm{d}x \, x H(x, \xi, t) & = & M_2(t) + \frac{4}{5} \xi^2 d_1(t),   
\label{mellin}
\end{eqnarray}
where the GFF $d_1(t)$ encodes the distribution of shear forces on the quarks and the pressure 
distribution in the proton. Ideally, one would determine the integral by measuring GPD $H$ in the entire $x$ and $\xi$ space 
in a large range of $t$. 
Given the current state of the DVCS experiments, such an approach is impractical as the GPDs are not directly 
accessible in experiment in the full $x$-space, but only at $x = \pm \xi$. We therefore proceed with a
 more phenomenological approach and express the GPD $H(x, \xi, t)$ in terms of the Compton Form Factor 
 $\mathcal{H}(\xi, t)$ through the convolution integral defined as 
\begin{multline} 
$$\text{Re}{\mathcal H}(\xi,t) + i \text{Im}{\mathcal H}(\xi,t)  = \nonumber \\
\int_{0}^{1} dx \left[ \frac{1}{\xi-x-i\epsilon} -  \frac{1}{\xi+x-i\epsilon} \right] H(x,\xi,t),\\  \nonumber  
$$\end{multline}
\noindent where we have traded the real function of 3 parameters $H(x, \xi, t)$ with the complex function of 2 parameters  
$\text{Re} {\mathcal H(\xi,t)}$ and $\text{Im}{\mathcal H}(\xi,t)$, which can be related more directly to 
experimentally accessible observables.
\begin{figure}[ht]
\includegraphics[width=0.9\columnwidth]{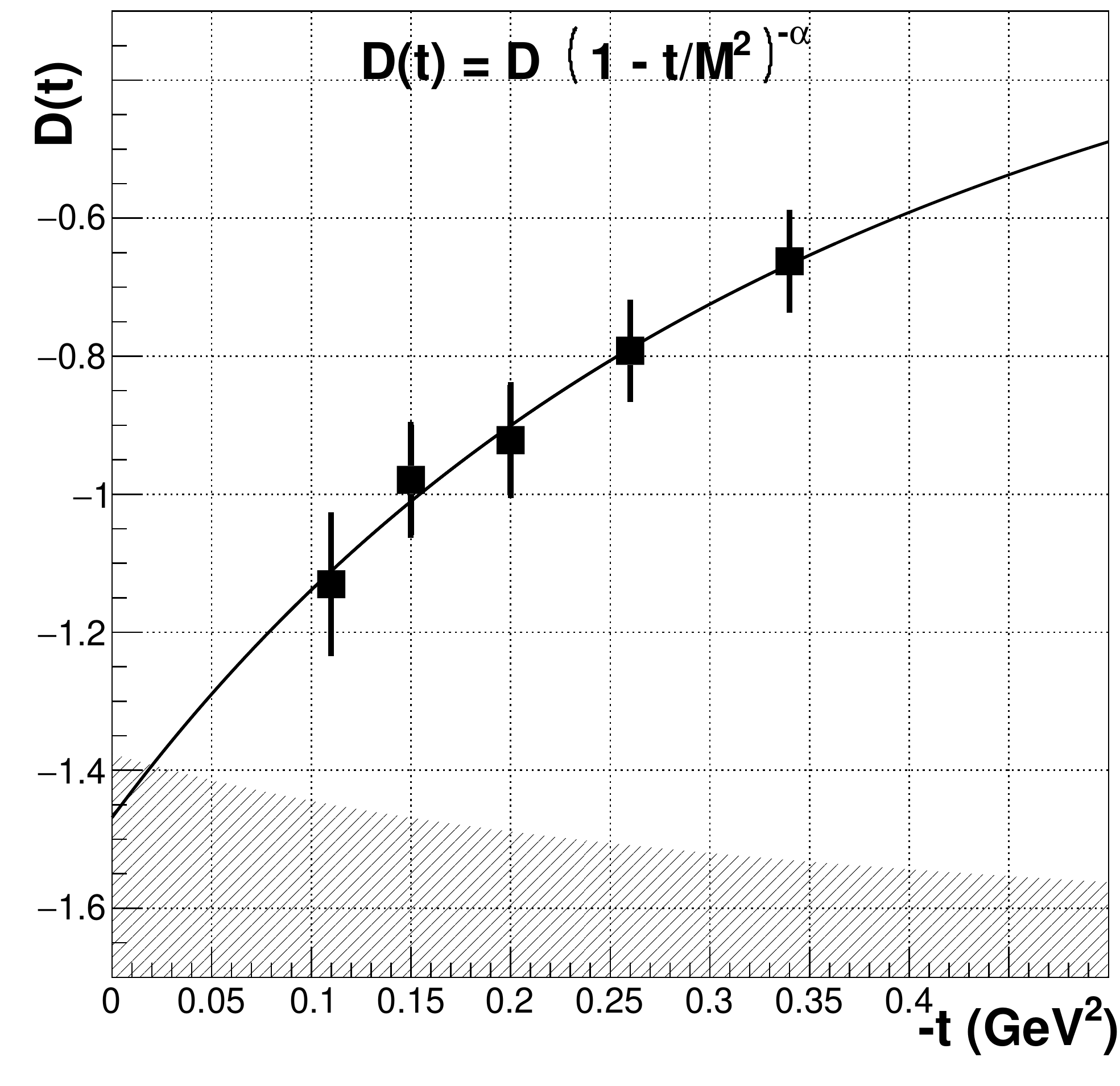}
\caption{The form factor $D(t)$ as determined in the fit to the DVCS data. The hatched area represents the 
magnitude of the estimated systematic uncertainties.}
\label{Dt}
\end{figure}

This analysis uses experimental data of DVCS with polarized electron beams that have 
been measured using the CLAS detector at Jefferson Laboratory in limited kinematics of $\xi$ and $t$. 
The imaginary part and the real part of $\cal{H}(\xi, \it{t})$ are extracted in local fits to the CLAS  beam-spin asymmetry 
data~\cite{Girod:2007aa} and to the cross section data~\cite{Jo:2015ema}, respectively. 
In order to cover the full $\xi$ range we use parameterizations~\cite{Mueller:2013caa,Kumericki:2016ehc} that were obtained 
from global fits to world data. The global and local fits show good agreement in $\xi$ and $t$ kinematics where they overlap. 

${\rm Im}\cal{H}(\xi, \it{t})$ and ${\rm Re}\cal{H}(\xi, \it{t})$ are related through a dispersion relation at fixed values of $t$.
 \begin{multline}
$$\text{Re}{\mathcal H}(\xi,t) \stackrel{\rm lo}{=} \\ D(t) \nonumber
+ {\mathcal P}\int_{0}^{1} dx  \left[\frac{1}{\xi-x}-\frac{1}{\xi+x}\right] \text{Im}{\mathcal H}(\xi,t) \nonumber
$$
\end{multline}

In this approach, we parameterized both ${\rm Im}\cal{H}(\xi, \it{t})$ and $D(t)$ and fitted these parameters directly to the data, 
with the dispersion relation included in the fit. Therefore we only have a few parameters adjusted to the entire available 
kinematic space. The parameterization provides strong constraints, which is currently the only way to obtain physical results given the status of the experiments. The systematic uncertainties inherent in this approach are carefully evaluated and are included in the results. 

From the global fit we determine $D(t)$ for each value of $\xi$. $D(t)$ is directly related 
to $d_1(t)$ in eq.(\ref{mellin}) and to the mechanical properties of the proton~\cite{Diehl:2007jb,Anikin:2007tx,Pasquini:2014vua}.  
The GFF $d_1(t)$ appears as the first coefficient in the expansion of the $D(t)$-term in terms of 
Gegenbauer polynomials in $\xi$. Details of the global fit and the extraction of $D(t)$ from the experimental 
data are discussed in a forthcoming article~\cite{Burkert2021}.

In Fig.~\ref{Dt} the results of the $D(t)$ form factor extraction are displayed, and the fit to the multipole form:
\begin{eqnarray}
 D(t) &=& D \bigg[1 + \frac{-t}{M^2}\bigg]^{-\alpha}, \\  \nonumber
 \label{quadrupole}
 \end{eqnarray} 
where $D$, $\alpha$ and $M^2$ are the fit parameters.
\begin{figure}[h]
\includegraphics[width=1.0\columnwidth]{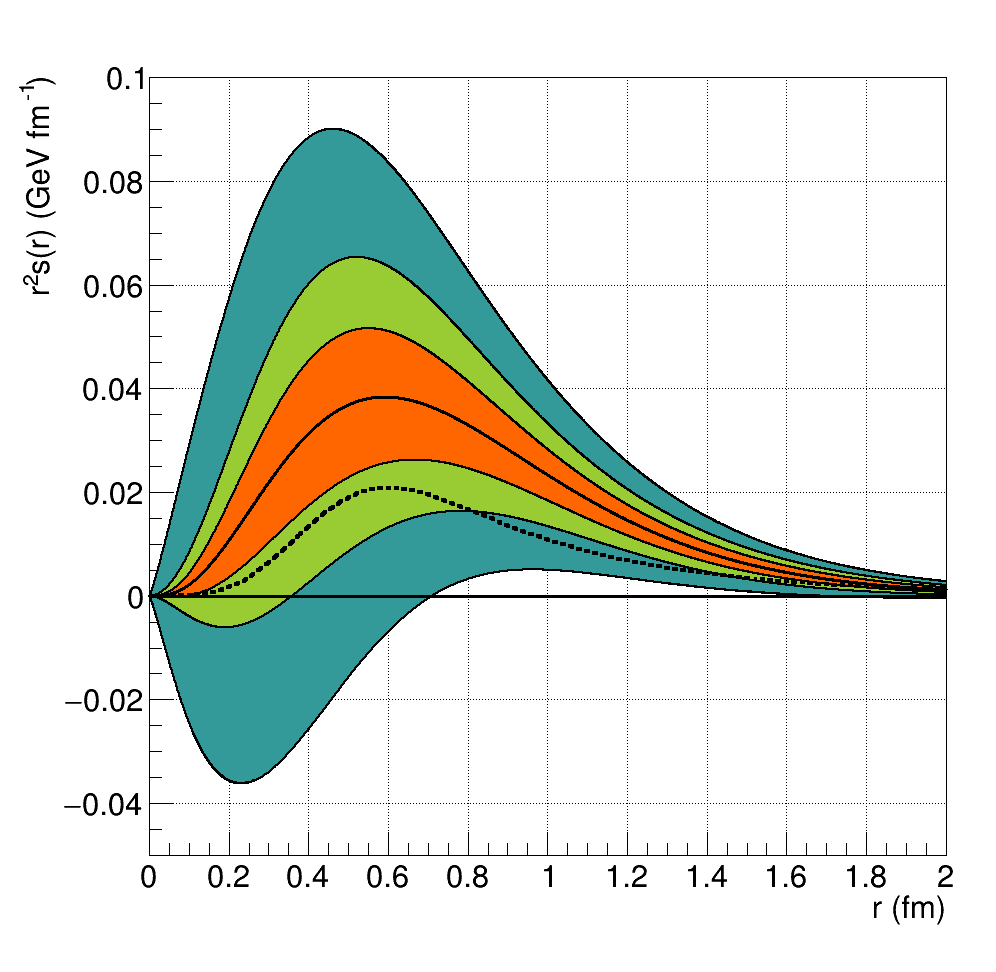}
\caption{The distribution of the shear forces $s(r)$ in the proton. The middle solid line represents the fit result. The 
outer blue-shaded area marks the range of uncertainties when only data prior to the CLAS data are included. The 
middle (light-green) areas are based on the CLAS data, and the inner (red) area represent projections when expected 
results from the ongoing and planned experiments are included in the fits. The widths of the bands are 
dominated by systematic uncertainties in parameterizations used in the integration. 
The dashed black curve is a model prediction ~\cite{Polyakov:2018zvc}.}
\label{shear-stress-1D}
\end{figure} 
Our fits result in the following parameters: 
\begin{eqnarray}
D &=&-1.47\pm 0.06 \pm 0.14 \label{F1} 
\\ 
M^2 &=&+1.02\pm  0.13 \pm 0.21  {\rm ~GeV^2}  \label{F2}\\ 
 \alpha &=&+2.76 \pm  0.23 \pm 0.48~ \label{F3} , 
 \\ \nonumber
\end{eqnarray} 
 \noindent where the first error is the fit uncertainty, and the second error is due to the systematic uncertainties.
  \begin{figure}[t!]
\includegraphics[width=0.95\columnwidth]{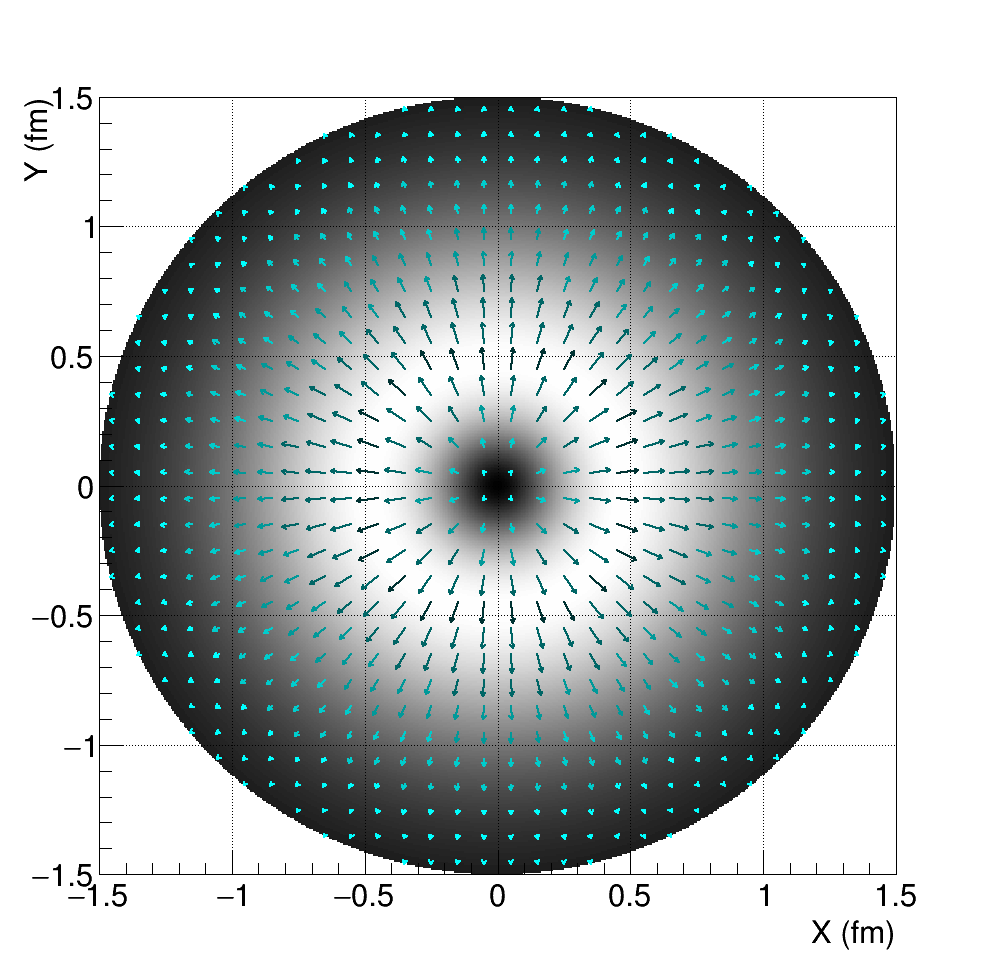}
\includegraphics[width=0.95\columnwidth]{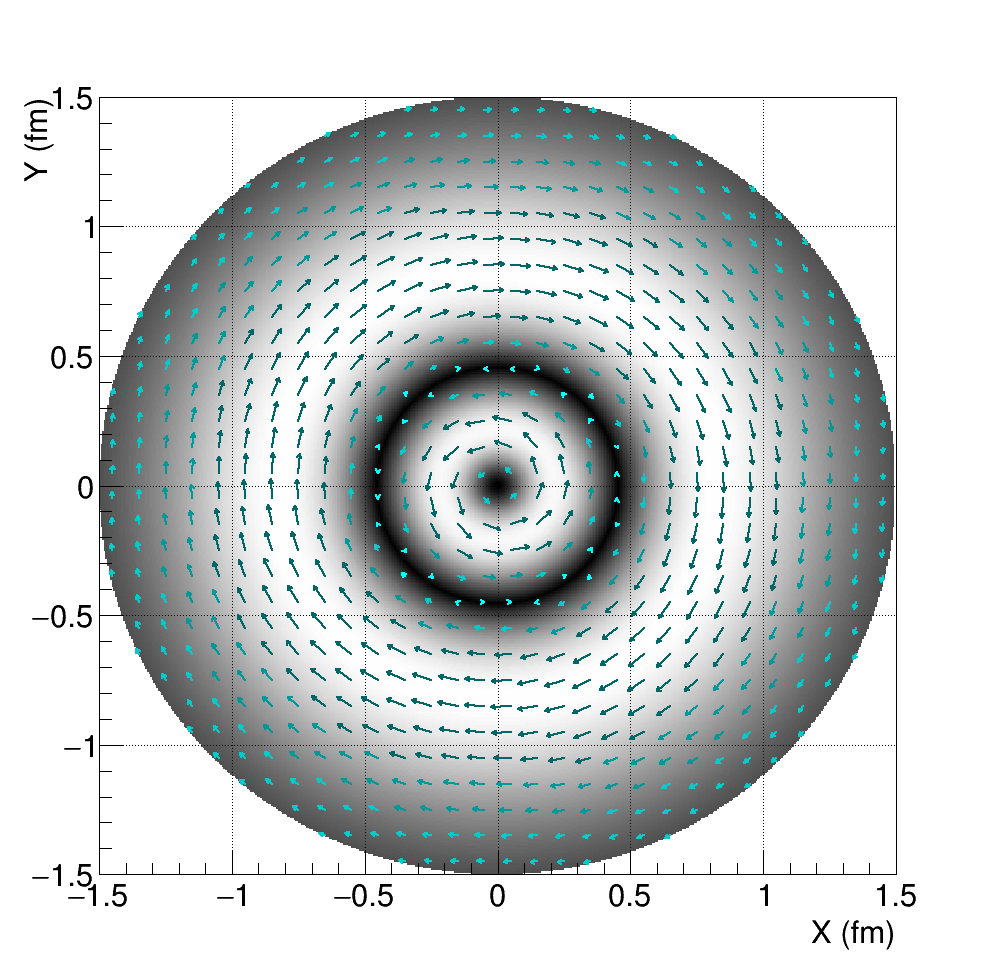}
\caption{\footnotesize 2D display of the the distribution of forces on quarks in the proton versus the distance from the proton center. Top: Normal force distribution. Bottom: Tangential or shear force distribution. The light gray shading and longer arrows indicate areas of stronger forces, the dark shading and shorter arrows indicate areas of weaker forces.  The shear  forces exhibit a node near a distance $r \approx  0.45$fm from the center, where they also reverse direction as indicated by the direction and lengths of the colored arrows.}
\label{shear_stress}
\end{figure}  

Approaches that do not rely on parameterizations or other constraints, such as techniques used in \cite{Kumericki:2019ddg,Dutrieux:2021nlz}, 
have so far not been successful in determining mechanical properties of the proton from data.

Next we relate $D(t)$ to the static energy-momentum tensor (EMT) components.  
The shear forces $s(r)$ and the pressure distribution $p(r)$ are related to the $ij$ components 
of the static energy momentum tensor as:  
\begin{eqnarray}
 T^{ij}({\bf r}) = \left({r^ir^j\over r^2} - {1\over 3}\delta^{ij}\right)s(r) + \delta^{ij}p(r)
\end{eqnarray} 
The shear forces are defined separately for quarks and gluons. Therefore by using $D(t)$ from the 
DVCS process the shear forces $s(r)$ and normal pressure $p(r)$ are separately determined for the quark content of the proton.
In the Breit-frame they are expressed through the integral~\cite{Polyakov:2018zvc}: 
\begin{eqnarray}
 s(r) &=& - {1\over 4m} r {d \over dr} {1 \over r} {d \over dr} {\tilde{D}}(r) \\ 
 \nonumber
 \\ 
 p(r) &=& {1\over 6m} {1\over r^2} {d\over dr} {r^2}  {d\over dr} {\tilde{D}}(r) \\ \nonumber
\\ \nonumber
\tilde{D}(r) &=& \int{d^3{\bf\Delta} \over (2\pi)^3}}e^{-i{\bf\Delta} r}{D(-\bf{\Delta^2)}. \nonumber
\end{eqnarray}  

\noindent 
Where $\bf{\Delta}$ is the 3-momentum transfer to the proton. We neglect contributions from the GFF $\bar{c}$~\cite{Polyakov:2018zvc} 
which would cancel with the gluons in the total Energy Momentum Tensor. Further extensions and discussions of other reference 
frames are included in~\cite{Lorce:2018egm}. 

When carrying out the integral over the entire kinematical range, we extract $s(r)$ and $p(r)$. 
Fig.~\ref{shear-stress-1D} shows the results for shear stress.  
This result represents the first effort in determining the shear forces in the proton using 
the experimental DVCS process and its relation to the GFFs. While our results still have 
significant systematic uncertainties due to the limited kinematic range covered in the DVCS data, 
they lead to interesting conclusions. The shear forces in the proton are tangential to the surface, and they 
change their direction near 0.45~fm radial distance from the center.  The maximum shear force of 
$40 \pm 20 $~MeVfm$^{-1}$ occurs near 0.55~fm from the proton center, 
indicating the location where confinement forces may be dominant. On the macroscopic scale of the 
earth surface, this force corresponds to the weight of a mass of 650~kg. 
A 2-dimensional display of both the normal and shear forces is presented in Fig.~\ref{shear_stress}. 
We want to stress that the arrows represent the forces acting along the orientation of the surface. So if one 
changes the direction of the normal to the surface one also changes the arrow direction, so that pressure 
acts equally on both sides of a surface immersed in the system.  
 
Shear forces in the proton have been computed in models~\cite{Goeke:2007fp,Polyakov:2018zvc,Lorce:2018egm,Shanahan:2018nnv,Freese:2021czn,Panteleeva:2021iip}. 
New experiments are currently underway~\cite{CLAS12:RGK} to measure the DVCS process with much 
extended kinematical coverage and higher statistical precision. This should lead to significantly reduced 
systematic uncertainties in the extraction of the gravitational properties of protons and neutrons, including their 
shear forces, as indicated in Fig.~\ref{shear-stress-1D} as the innermost (red) shaded area.      

\section{Acknowledgments} 
We are grateful for useful comments by Cedric Lorce and Maxim Polyakov. 
The material discussed in the paper is based on research supported by the US Department of Energy, Office of Science, 
Office of Nuclear Physics under contract DE-AC05-06OR23177.


\begin{thebibliography}{99}

\bibitem{Kobzarev:1962wt} 
  I.~Y.~Kobzarev and L.~B.~Okun,
  ``Gravitational Interaction Of Fermions,''
  Zh.\ Eksp.\ Teor.\ Fiz.\  {\bf 43}, 1904 (1962)
  [Sov.\ Phys.\ JETP {\bf 16}, 1343 (1963)].

\bibitem{Pagels:1966zza} 
  Pagels,~H.
 ``Energy-Momentum Structure Form Factors of Particles'',
  Phys.\ Rev.\  {\bf 144}, 1250 (1966)
 doi:10.1103/PhysRev.144.1250.

\bibitem{Ji:1996ek} 
  Ji,~X.~D.
 ``Gauge-Invariant Decomposition of Nucleon Spin'',
  Phys.\ Rev.\ Lett.\  {\bf 78}, 610 (1997)
 doi:10.1103/PhysRevLett.78.610
 [hep-ph/9603249].

\bibitem{Ji:1996nm} 
  Ji,~X.~D.
  ``Deeply virtual Compton scattering'',
  Phys.\ Rev.\ D {\bf 55}, 7114 (1997)
  doi:10.1103/PhysRevD.55.7114
  [hep-ph/9609381].

\bibitem{Radyushkin:1996nd} 
Radyushkin,~A.~V.
  ``Scaling limit of deeply virtual Compton scattering'',
  Phys.\ Lett.\ B {\bf 380}, 417 (1996)
 doi:10.1016/0370-2693(96)00528-X
  [hep-ph/9604317].     
   
\bibitem{Mueller:1994} 
  M\"uller,~D., Robaschik,~D., Geyer,~D., Dittes,~F.\-M., \& Ho\v{r}ej\v{s}i,~J.
  ``Wave functions, evolution equations and evolution kernels from light ray operators of QCD'',
  Fortsch.\ Phys.\  {\bf 42}, 101 (1994)
   doi:10.1002/prop.2190420202
  [hep-ph/9812448]. 

\bibitem{Polyakov:2002yz} 
 Polyakov,~M.~V. 
 ``Generalized parton distributions and strong forces inside nucleons and nuclei'',
  Phys.\ Lett.\ B {\bf 555}, 57 (2003)
 doi:10.1016/S0370-2693(03)00036-4
[hep-ph/0210165]. 
	
\bibitem{Burkert:2018bqq} 
V.~D.~Burkert, L.~Elouadrhiri and F.~X.~Girod,
``The pressure distribution inside the proton,''
Nature {\bf 557}, no. 7705, 396 (2018).
doi:10.1038/s41586-018-0060-z

\bibitem{Girod:2007aa} 
Girod,~F.~X {\it et al.} [CLAS Collaboration],
``Measurement of Deeply virtual Compton scattering beam-spin asymmetries'',
Phys.\ Rev.\ Lett.\  {\bf 100}, 162002 (2008)
doi:10.1103/PhysRevLett.100.162002
[arXiv:0711.4805 [hep-ex]].
	
\bibitem{Jo:2015ema} 
Jo,~H.~S. {\it et al.} [CLAS Collaboration],
``Cross sections for the exclusive photon electroproduction on the proton and Generalized Parton Distributions'',
Phys.\ Rev.\ Lett.\  {\bf 115}, no. 21, 212003 (2015)
doi:10.1103/PhysRevLett.115.212003
[arXiv:1504.02009 [hpe-ex]].

\bibitem{Polyakov:2018zvc} 
M.~V.~Polyakov and P.~Schweitzer,
``Forces inside hadrons: pressure, surface tension, mechanical radius, and all that,''
Int.\ J.\ Mod.\ Phys.\ A {\bf 33}, no. 26, 1830025 (2018)
doi:10.1142/S0217751X18300259
[arXiv:1805.06596 [hep-ph]].

\bibitem{Lorce:2018egm}
C.~Lorc\'e, H.~Moutarde and A.~P.~Trawi\'nski,
``Revisiting the mechanical properties of the nucleon,''
Eur. Phys. J. C \textbf{79}, no.1, 89 (2019)
doi:10.1140/epjc/s10052-019-6572-3
[arXiv:1810.09837 [hep-ph]].

\bibitem{Polyakov:1999gs}
M.~V.~Polyakov and C.~Weiss,
``Skewed and double distributions in pion and nucleon,''
Phys. Rev. D \textbf{60}, 114017 (1999)
doi:10.1103/PhysRevD.60.114017
[arXiv:hep-ph/9902451 [hep-ph]]

\bibitem{Goeke:2007fp} 
  K.~Goeke, J.~Grabis, J.~Ossmann, M.~V.~Polyakov, P.~Schweitzer, A.~Silva and D.~Urbano,
  ``Nucleon form-factors of the energy momentum tensor in the chiral quark-soliton model,''
  Phys.\ Rev.\ D {\bf 75}, 094021 (2007)
  doi:10.1103/PhysRevD.75.094021
  [hep-ph/0702030].

\bibitem{Diehl:2007jb} 
 Diehl,~M., \& Ivanov,~D.~Y.
 ``Dispersion representations for hard exclusive processes: beyond the Born approximation'',
  Eur.\ Phys.\ J.\ C {\bf 52}, 919 (2007)
  doi:10.1140/epjc/s10052-007-0401-9
  [arXiv:0707.0351 [hep-ph]].

\bibitem{Kumericki:2016ehc} 
  Kumericki,~K., Liuti,~S., \& Moutarde,~H.
  ``GPD phenomenology and DVCS fitting : Entering the high-precision era'',
  Eur.\ Phys.\ J.\ A {\bf 52}, no. 6, 157 (2016)
  doi:10.1140/epja/i2016-16157-3
  [arXiv:1602.02763 [hep-ph]].

\bibitem{Mueller:2013caa}
  M\"uller,~D., Lautenschlager,~T.,  Passek-Kumericki,~K., \& Schaefer,~A.
  ``Towards a fitting procedure to deeply virtual meson production - the next-to-leading order case'',
  Nucl.\ Phys.\ B {\bf 884} (2014) 438
 doi:10.1016/j.nuclphysb.2014.04.012
  [arXiv:1310.5394 [hep-ph]].

\bibitem{Anikin:2007tx} 
  Anikin,~I.~V., \& Teryaev,~O.~V.
``Dispersion relations and QCD factorization in hard reactions'',
  Fizika B {\bf 17}, 151 (2008)
  [arXiv:0710.4211 [hep-ph]].

\bibitem{Pasquini:2014vua} 
  Pasquini,~B., Polyakov,~M.~V., \& Vanderhaeghen,~M.
  ``Dispersive evaluation of the D-term form factor in deeply virtual Compton scattering'',
  Phys.\ Lett.\ B {\bf 739}, 133 (2014)
  doi:10.1016/j.physletb.2014.10.047
  [arXiv:1407.5960 [hep-ph]].

\bibitem{Shanahan:2018nnv}
P.~E.~Shanahan and W.~Detmold,
``Pressure Distribution and Shear Forces inside the Proton,''
Phys. Rev. Lett. \textbf{122}, no.7, 072003 (2019)
doi:10.1103/PhysRevLett.122.072003
[arXiv:1810.07589 [nucl-th]].

\bibitem{Azizi:2019ytx}
K.~Azizi and U.~\"Ozdem,
``Nucleon\textquoteright{}s energy\textendash{}momentum tensor form factors in light-cone QCD,''
Eur. Phys. J. C \textbf{80}, no.2, 104 (2020)
doi:10.1140/epjc/s10052-020-7676-5
[arXiv:1908.06143 [hep-ph]]
  
  \bibitem{Anikin:2019kwi} 
  I.~V.~Anikin, 
  ``Gravitational form factors within light-cone sum rules at leading order,''
  Phys.\ Rev.\ D {\bf 99}, no. 9, 094026 (2019)
  doi:10.1103/PhysRevD.99.094026
  [arXiv:1902.00094 [hep-ph]].

\bibitem{Kumericki:2019ddg}
K.~Kumeri\v{c}ki,
``Measurability of pressure inside the proton,''
Nature \textbf{570}, no.7759, E1-E2 (2019)
doi:10.1038/s41586-019-1211-6

\bibitem{Dutrieux:2021nlz}
H.~Dutrieux, C.~Lorc\'e, H.~Moutarde, P.~Sznajder, A.~Trawi\'nski and J.~Wagner,
``Phenomenological assessment of proton mechanical properties from deeply virtual Compton scattering,''
[arXiv:2101.03855 [hep-ph]].
   
\bibitem{Burkert2021} V. Burkert, L. Elouadrhiri, and F.X. Girod,  paper in preparation. 

\bibitem{Freese:2021czn} 
A.~Freese and G.~A.~Miller,  
``Forces within hadrons on the light front,'' 
[arXiv:2102.01683 [hep-ph]]. 

\bibitem{Panteleeva:2021iip} 
J.~Y.~Panteleeva and M.~V.~Polyakov, 
 ``Forces inside the nucleon on the light front from 3D Breit frame force distributions: Abel tomography case,''
  [arXiv:2102.10902 [hep-ph]]. 

\bibitem{Hagler:2007xi} 
  P.~Hagler {\it et al.} [LHPC Collaboration],
  ``Nucleon Generalized Parton Distributions from Full Lattice QCD,''
  Phys.\ Rev.\ D {\bf 77}, 094502 (2008)
  doi:10.1103/PhysRevD.77.094502
  [arXiv:0705.4295 [hep-lat]].
  
\bibitem{CLAS12:RGK}
  Elouadrhiri,~L. {\it et al.} [CLAS Collaboration],
  `Deeply Virtual Compton Scattering with CLAS12 at 6.6 GeV and 8.8 GeV'',
  E12-16-010B approved by Jefferson Lab PAC44

\end{thebibliography}
\end{document}